\documentstyle[preprint,aps,epsfig]{revtex}
\tightenlines
\begin{document}
\preprint{\vbox{\hbox{IFT--P.037/2002} \hbox{IFUSP-1548/2002}}}
\title{Radion and Higgs Signals in Peripheral Heavy Ion Collisions
at the LHC.}
\author{ S.\ M.\ Lietti$^1$ and C.\ G.\ Rold\~ao$^2$}
\address{$^1$Instituto de F\'{\i}sica da USP, \\
         C.P. 66.318, S\~ao Paulo, SP 05389-970, Brazil.\\
         $^2$Instituto de F\'{\i}sica Te\'orica,
         Universidade  Estadual Paulista, \\
         Rua Pamplona 145, CEP 01405-900 S\~ao Paulo, Brazil.}
\date{\today}
\maketitle
\widetext
\begin{abstract}
We investigate the sensitivity of the heavy ion mode of the LHC to
Higgs boson and Radion production via photon-photon fusion through 
the analysis of the processes $\gamma \gamma \to \gamma \gamma$, 
$\gamma \gamma \to b \bar{b}$, and $\gamma \gamma \to g g$ in peripheral 
heavy ion collisions. We suggest cuts to improve the Higgs and Radion 
signal over standard model background ratio and determine the capability 
of LHC to detect these particles production.
\end{abstract}


\section{Introduction}
\label{sec:int}

The standard model (SM) has been very successful in accounting
for almost all experimental data. The Higgs boson is the only
particle in the SM that has not yet been confirmed experimentally. 
It is responsible for the mass generation of fermions and gauge bosons. 
The search for the Higgs boson is the main priority in high energy 
experiments and hints of its existence may have been already seen 
at LEP\cite{Higgs} at around $m_H \sim 115$ GeV. Nevertheless, the SM 
can only be a low energy limit of a more fundamental theory because it 
cannot explain a number of theoretical issues, one of which is the gauge
hierarchy problem between the only two known scales in particle
physics -- the weak and Planck scales. Recent advances in string
theories have revolutionized our perspectives and understanding
of the problems, namely, the Planck, grand unification, and
string scales can be brought down to a TeV range with the help of
extra dimensions, compactified or not.   Arkani-Hamed {\it et
al.} \cite{add} proposed that using compactified dimensions of
large size (as large as mm) can bring the Planck scale down to
TeV range. Randall and Sundrum \cite{RS} proposed a
5-dimensional space-time model with a nonfactorizable metric to
solve the hierarchy problem. The Randall-Sundrum model (RSM) has a 
four-dimensional massless scalar, the modulus or Radion. 
The most important ingredients of the above model are the 
required size of the Radion field such that it generates the desired 
weak scale from the scale $M$ ($\approx$ Planck scale) and the
stabilization of the Radion field at this value.
A stabilization mechanism was proposed by Goldberger and Wise
\cite{GW}. As a consequence of this stabilization, the mass of the 
Radion is of order of $O(\rm TeV)$ and the strength
of coupling to the SM fields is of order of $O(1/{\rm TeV})$.
Therefore, the detection of this Radion will be the first
signature of the RSM and the stabilization mechanism by Goldberger and
Wise.  

Higgs and Radion can be produced in various types of accelerators. Several
papers have been published in order to study the possibility of detection of
the Higgs particle in $e^+ e^-$, $\mu^+ \mu^-$, $p \bar{p}$, $p p$ and
$\gamma \gamma$ colliders \cite{Higgs_search}. Recently, the phenomenology
of the Radion particle has been also studied for  $e^+ e^-$, $p p$ and
$\gamma \gamma$ colliders \cite{kingman}.
In this paper we explore the possibility of an intermediate-mass
Higgs boson or Radion scalar be produced  in peripheral heavy ion
collisions through photon-photon interactions \cite{baur,baur2}.
The reason to choose photon-photon fusion in  peripheral heavy
ion collisions resides in the fact that the production mode is
free of any problem caused by strong interactions of the initial
state, which make these processes cleaner than pomeron-pomeron or
pomeron-photon fusions. In the context of the SM, the Higgs boson
has been explored in detail in the literature
\cite{papa,cahn}, with the general conclusion that the
chances of finding the SM Higgs in the photon-photon
case are marginal. On the other hand, a study of Radion production
in peripheral heavy ion collisions has not yet been made.

The Higgs couplings considered in this paper are given by the usual SM
lagrangian while the Radion effects can be described by effective operators
involving the spectrum of the SM and the Radion scalar field.
The Radion couplings to the SM particles are similar to the Higgs
couplings to the same particles, except from a factor involving the Higgs and
the Radion vacuum expectation values (vev's), as can be seen in 
Section \ref{sec:eff}.
In Section \ref{sec:sim} we present the strategy to evaluate
photon-photon fusion processes in peripheral heavy ion collisions
and in Section \ref{sec:res}
we explore the capabilities of peripheral heavy ion
collisions in detecting Higgs and Radion productions by
analyzing the processes $\gamma \gamma \to \gamma\gamma$, $b \bar{b}$, 
and $gg$. After simulating the signal and
background, we find optimal cuts to maximize their ratio. 
We show how to use the  invariant mass spectra of the final state 
$\gamma\gamma$, $b \bar{b}$, and $gg$ pairs in order to improve the 
SM Higgs boson and RMS Radion signals. Finally, in Section \ref{sec:con} 
we draw our final conclusions.

\section{Effective Lagrangian for the Radion Couplings}
\label{sec:eff}

In order to describe the interactions of the RSM Radion 
with the SM particles, we follow the notation of Ref.\ \cite{kingman}. 
These interactions are model-independent and are governed by 
4-dimensional general covariance, and thus given by the following Lagrangian
\begin{equation}
\label{T}
{\cal L}_{\rm int} = \frac{R}{\Lambda_R} \; T^\mu_\mu ({\rm SM}) \;,
\end{equation}
where $\Lambda_R= \langle R \rangle$ is of order TeV, and $T_\mu^\mu$
is the trace of SM energy-momentum tensor, which is given by
\begin{equation}
T^\mu_\mu ({\rm SM}) = \sum_f m_f \bar f f - 2 m_W^2 W_\mu^+ W^{-\mu}
-m_Z^2 Z_\mu Z^\mu + (2m_H^2 H^2 - \partial_\mu H \partial^\mu H  ) + ... \;,
\end{equation}
where $...$ denotes higher order terms.
The couplings of the Radion with fermions and $W$, $Z$ and Higgs
bosons are given in Eq. (\ref{T}). Note that the couplings of the Radion with
fermions, $W$, and $Z$ are similar to the couplings of the Higgs to these
particles, the only difference resides in the coupling constants
where $v$, the vev of the Higgs field, is replaced by  $\Lambda_R$.

The coupling of the Radion to a pair of gluons
(photons) is given by contributions from 1-loop diagrams with the
top-quark (top-quark and $W$) in the loop, similar to the Higgs boson
couplings to the same pair.
However, for the Radion case, there is another contribution coming
from the trace anomaly for gauge fields, that is given by
\begin{equation}
T^\mu_\mu({\rm SM})^{\rm anom} = \sum_a \frac{\beta_a (g_a)}{2g_a}
F_{\mu\nu}^a F^{a \mu\nu} \;.
\end{equation}

For the coupling of the Radion to a pair of gluons,
$\beta_{\rm QCD}/2g_s = -(\alpha_s/8\pi) b_{\rm QCD}$, where
$b_{\rm QCD} = 11 - 2 n_f/3$ with $n_f=6$.  Thus,
the effective coupling of $R g(p_{1,\mu,a}) g(p_{2,\nu,b})$,
including the 1-loop diagrams of top-quark
and the trace anomaly contributions is given by
\begin{equation}
\frac{i \delta_{ab}\alpha_s}{2\pi \Lambda_R}
\left[ b_{\rm QCD} + y_t ( 1+ (1-y_t)f(y_t) ) \right ]
\left( p_1 \cdot p_2 g_{\mu \nu} - p_{2_\mu} p_{1_\nu} \right ) \;,
\end{equation} where $y_t= 4 m_t^2/2p_1 \cdot p_2$.

The effective coupling of $R \gamma(p_{1,\mu})\gamma(p_{2,\nu})$,
including the 1-loop diagrams of the top-quark and $W$ boson, and the trace
anomaly contributions  is given by
\begin{eqnarray}
\frac{i \alpha_{\rm em}}{2\pi \Lambda_R}
\left[ b_2 + b_Y - (2+3y_W +3y_W (2-y_W)f(y_W) ) + \frac{8}{3} \,
y_t ( 1+ (1-y_t)f(y_t) ) \right ]
\nonumber\\
\times \left( p_1 \cdot p_2 g_{\mu \nu} - p_{2_\mu} p_{1_\nu} \right ) \;,
\label{r_photon_photon}
\end{eqnarray}
where $y_i= 4 m_i^2/2p_1 \cdot p_2$, $b_2=19/6$ and $b_Y=-41/6$.
In the above, the function $f(z)$ is given by
\begin{displaymath}
f(z) = \left \{ \begin{array}{cr}
\left[ \sin^{-1} \left(\frac{1}{\sqrt{z}} \right ) \right ]^2\;, & z \ge 1 \\
-\frac{1}{4} \left[ \log \frac{1+\sqrt{1-z}}{1-\sqrt{1-z}} - i \pi \right ]^2
\;, & z <1
\end{array}
\right . \;.
\end{displaymath}

Equations (\ref{T}--\ref{r_photon_photon}) give
all necessary couplings to perform calculations on decays and
production of the Radion. In order to perform calculations on the 
decays and production of the Higgs, we consider its SM couplings, widely
discussed in the literature.

\section{Simulations}
\label{sec:sim}

In order to perform the Monte Carlo analysis, we have employed the
package MadGraph \cite{Madgraph} coupled to HELAS \cite{helas}.
Special subroutines were constructed for the anomalous contribution which
enable us to take into account all interference effects between
the QED and the anomalous amplitudes. The phase space integration
was performed by VEGAS \cite{vegas}.

The photon distribution in the nucleus can be described using the
equivalent-photon or Weizs\"{a}cker-Williams approximation in the impact
parameter space. Denoting the photon distribution function in a nucleus by
$F(x)$, which represents the number of
photons carrying a fraction between $x$ and $x+dx$ of the
total momentum of a nucleus of charge $Ze$, we can define
the two-photon luminosity through
\begin{equation}
\frac{dL}{d\tau} = \int ^1 _\tau \frac{dx}{x} F(x) F(\tau/x),
\end{equation}

\noindent
where $\tau = {\hat s}/s$, $\hat s$ is the square of the center
of mass (c.m.s.) system energy of the two photons and $s$ of the
ion-ion system. The total cross section
$ AA \rightarrow AA \gamma \gamma \rightarrow AA X$, where
$X$ are the particles produced by the $\gamma \gamma$ process, is
\begin{equation}
\sigma (s) = \int d\tau \frac{dL}{d\tau} \hat \sigma(\hat s),
\label{sigfoton}
\end{equation}
where $ \hat \sigma(\hat s)$ is the cross-section of the subprocess
$\gamma \gamma \rightarrow X$.

We choose to use the conservative and more realistic photon
distribution of Cahn and Jackson~\cite{cahn}, including a
prescription proposed by Baur~\cite{baur2} for realistic
peripheral collisions, where we must enforce that the minimum
impact parameter ($b_{min}$) should be larger than $R_1 + R_2$,
where $R_i$ is the nuclear radius of the ion $i$. A useful fit
for the two-photon luminosity is:
\begin{equation}
\frac{dL}{d\tau}=\left(\frac{Z^2 \alpha}{\pi}\right)^2 \frac{16}{3\tau}
\xi (z),
\label{e3}
\end{equation}
where $z=2MR\sqrt{\tau}$, $M$ is the nucleus mass, $R$ its radius and
$\xi(z)$ is given by
\begin{equation}
\xi(z)=\sum_{i=1}^{3} A_{i} e^{-b_{i}z},
\label{e4}
\end{equation}
which is a fit resulting from the numerical integration of the photon
distribution, accurate to $2\% $ or better for $0.05<z<5.0$, and where
$A_{1}=1.909$, $A_{2}=12.35$, $A_{3}=46.28$, $b_{1}=2.566$,
$b_{2}=4.948$, and $b_{3}=15.21$. For $z<0.05$ we use the expression (see
Ref.~\cite{cahn})
\begin{equation}
\frac{dL}{d\tau}=\left(\frac{Z^2 \alpha}{\pi}\right)^2
\frac{16}{3\tau}\left[\ln{\left(\frac{1.234}{z}\right)}\right]^3 .
\label{e5}
\end{equation}
In this paper we consider electromagnetic processes of peripheral
Ar-Ar and Pb-Pb collisions in order to produce a Higgs and/or Radion
scalar via photon-photon fusion since the pomeron contributions are
negligible for subprocesses with center of mass energy close to the Higgs
mass. According to Ref. \cite{ion_review}, the total center of mass energy
for $^{40}_{18}$Ar ($^{208}_{82}$Pb) is equal to 7 (5.5) TeV/nucleon
and an average luminosity of $5.2 \times 10^{29} (4.2 \times
10^{26})$ cm$^{-2}$ s$^{-1}$,  which implies an effective
photon-photon luminosity for $m_{\gamma\gamma}=115$ GeV  equals
to $2 \times 10^{28} (8 \times 10^{26})$ cm$^{-2}$ s$^{-1}$
[0.63 (0.0025) pbarn$^{-1}$ year$^{-1}$] at LHC, as can be seen 
in Figure \ref{gg_lumi}, which was extracted from Ref. \cite{ion_review}. 
We will also consider the optimistic possibility of Ca-Ca collisions 
\cite{papageorgiu,our}, where the total center of mass energy 
for $^{40}_{20}$Ca is
equal to 7 TeV/nucleon and an average luminosity of $5 \times 10^{30}$
cm$^{-2}$ s$^{-1}$,  which implies an effective photon-photon luminosity for
$m_{\gamma\gamma}=115$ GeV  equals to $1.92 \times 10^{29}$ cm$^{-2}$ s$^{-1}$
(6 pbarn$^{-1}$ year$^{-1}$).

\section{Results}
\label{sec:res}

In our analyses, we computed the cross sections for
the Higgs and Radion production via photon-photon fusion in
peripheral heavy ion collisions at LHC, with the subsequent decay
of the Higgs and/or Radion into  $\gamma \gamma$, $b \bar{b}$ and
$gg$ pairs. The main sources of background for these processes
are the box diagram for the process $\gamma \gamma \to \gamma \gamma$,
the usual electromagnetic tree level diagrams for the process
$\gamma \gamma \to b \bar{b}$, and the box diagram $\gamma \gamma \to gg$
and the usual tree level diagrams $\gamma \gamma \to q \bar{q}$,
where $q=u,d,s,c$, for the process $\gamma \gamma \to  gg$.

We begin our analyses using similar cuts and efficiencies as the ones
ATLAS Collaboration \cite{atlas} applied in their studies of Higgs boson
searches. Our initial results are obtained imposing the following
acceptance set of cuts:
\begin{eqnarray}
p_{T}^{\gamma (b) [g]} > 25 \; \text{GeV} \;\;\; ,
\;\;\;\;\;\;
|\eta_{\gamma (b) [g]}| < 2.5 \;\;\; ,
\;\;\;\;\;\;
\Delta R_{\gamma \gamma (b\bar{b)[gg]}} > 0.4 \;\; ,
\label{cut1}
\end{eqnarray}
and taking into account an  efficiency for reconstruction and
identification of one photon of 84\%, an efficiency of reconstruction
for $H \to b \bar{b}$ of 90\% with a b-tagging of 60\% per each quark
$b$  \cite{atlas}, and finally an efficiency of reconstruction for
$H \to q \bar{q}$ or $gg$ of 80\%. Taking all these efficiencies into
account, the cross sections are evaluated with a total efficiency factor
of 70(32)[80]\% for the decay $H$ or $R \to \gamma \gamma (b
\bar{b})[gg]$.
The results are presented in Table \ref{acceptance_cut} for a Higgs
and Radion masses of 115 GeV, with $\Lambda_R = 4 v \approx 1$ TeV, 
in peripheral Ar-Ar and  Pb-Pb collisions at LHC. Results for
Ca-Ca collisions at LHC can also be obtained, according to Equation 
(\ref{e3}), by simply multiplying the results for Ar-Ar collisions 
by the factor $(\frac{Z_{Ca}}{Z_{Ar}})^4 = (\frac{20}{18})^4\approx1.524$.

In order to improve the Higgs and Radion signal over SM
background, i.e., all other Feynman diagrams that contribute to the
process considered, we have studied several kinematical distributions of the
final state particles.  Since the Higgs and Radion interactions occur mainly
when these particles are produced on-shell, the most promising one is the
invariant mass of the final particles. 

\begin{figure}
\protect
\centerline{\mbox{\epsfig{file=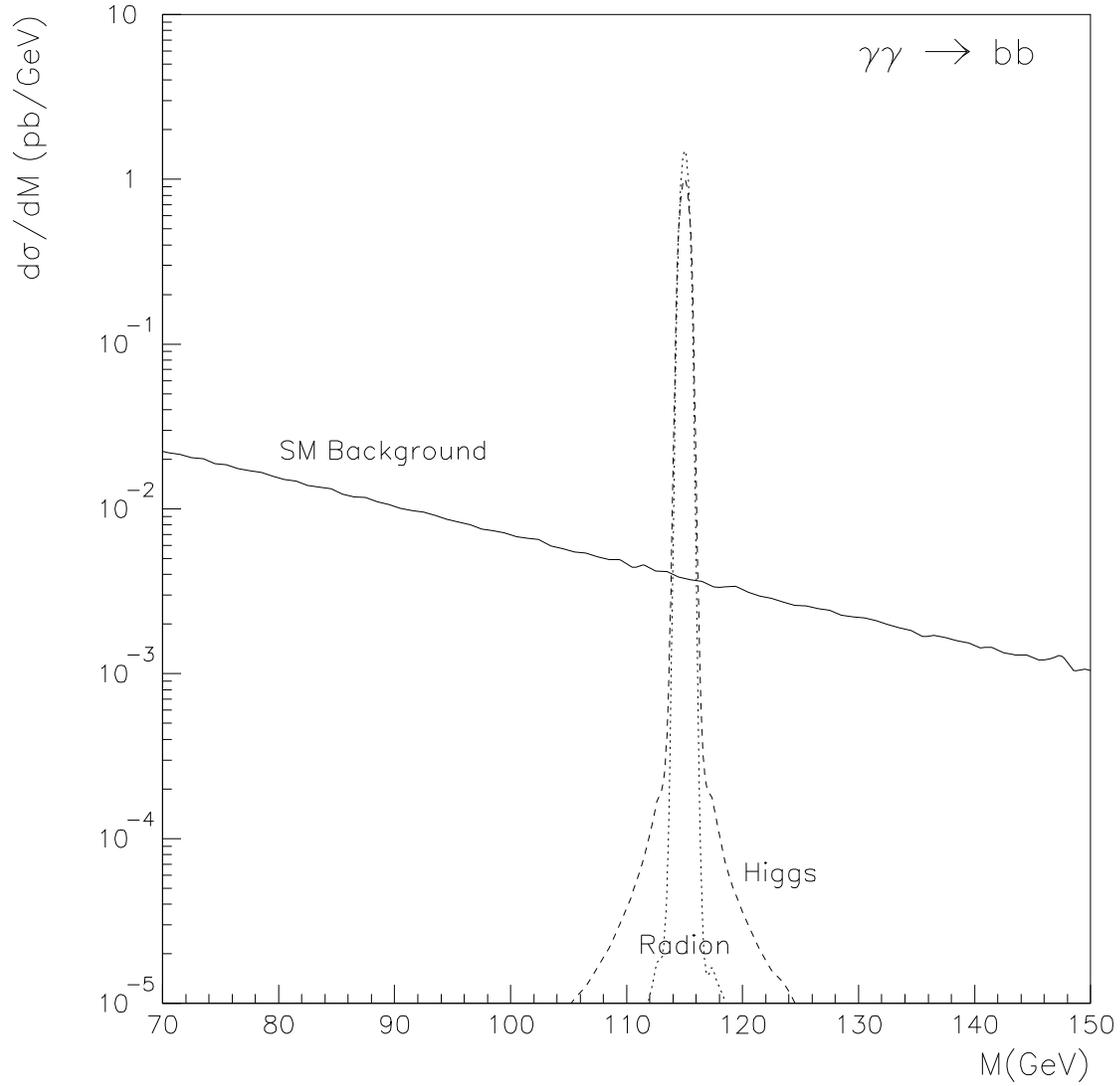,angle=0,width=1.0\textwidth}}}
\caption{Normalized invariant mass distribution of the $b\bar{b}$ pair
with a Higgs mass and a Radion mass equal to 115 GeV and 
$\lambda_R = 4v \approx 1$ TeV. 
The full line corresponds to the SM background discussed in the text 
while the dashed (dotted) line corresponds to the Higgs (Radion) contribution.}
\label{minv}
\end{figure}

The behavior of the normalized invariant mass distribution of the final  
state particles is plotted in Figure \ref{minv} for 
the process $\gamma \gamma \to b\bar{b}$ with a Higgs mass and a Radion 
mass equal to 115 GeV and $\lambda_R = 4v \approx 1$ TeV.
For instance, if we impose an additional cut of $|m_{b \bar{b}} - m_H|
< 15$ GeV in the process $\gamma \gamma \to b\bar{b}$, the value
for the SM background cross section in peripheral Ar-Ar collisions 
is reduced from 6.927 pb to 0.8486 pb, while the value for the 
Higgs (Radion) cross section ($\gamma \gamma \to H(R) \to b\bar{b}$)
is almost unaffected, varying from 0.1038(1.923$\times10^{-2}$) pb
to 0.1038(1.919$\times10^{-2}$) pb when the invariant mass cut is imposed.
Similar behavior is observed in the processes $\gamma \gamma \to
\gamma \gamma$ and $\gamma \gamma \to g g$, as can be seen in 
Table \ref{refined_cut}. 
Therefore we collected final states $\gamma \gamma$, $b \bar{b}$
and $gg$ events whose invariant masses fall in bins of size of 30 GeV 
around the Higgs (Radion) mass
\begin{eqnarray}
m_{H(R)} - 15\text{ GeV}
< m_{\gamma \gamma (b \bar{b}) [gg]}  < m_{H(R)} + 15\text{ GeV}
\;\;
\label{cut2}
\end{eqnarray}
in order to evaluate our results.

\begin{figure}
\protect
\centerline{\mbox{\epsfig{file=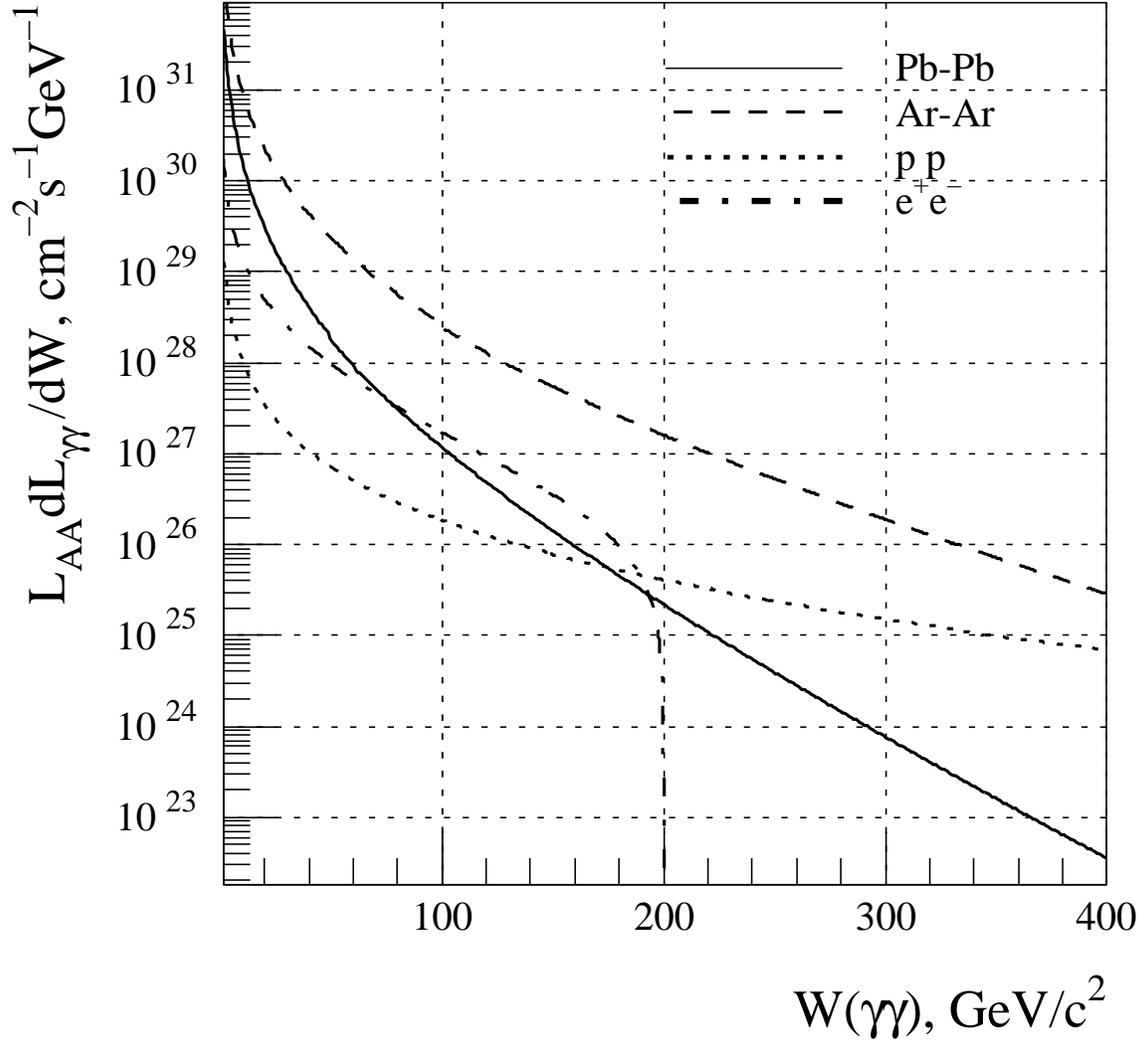,angle=0,width=1.0\textwidth}}}
\caption{Effective $\gamma \gamma$ luminosity at LHC for different
ion species and protons as well as the $e^+ e^-$ collider LEPII in terms
of the invariant mass of the pair of photons $W(\gamma \gamma)$.}
\label{gg_lumi}
\end{figure}

Considering the effective photon-photon luminosities given by Figure
\ref{gg_lumi} and Refs. \cite{ion_review,papageorgiu}, we 
note that the Ar-Ar(Ca-Ca) luminosity is $\approx$250(2500) times 
greater than the Pb-Pb luminosity. On the other hand, Table 
\ref{refined_cut} shows that the Pb-Pb cross sections are $\approx$30(20) 
times greater than the Ar-Ar(Ca-Ca) cross sections. Taking into account 
both luminosity and cross section behavior for each mode of the 
heavy ion LHC accelerator, one can realize that the total number of events 
in Ar-Ar(Ca-Ca) collisions  is $\approx$8(125) times greater than in 
Pb-Pb collisions, which shows that Pb-Pb collisions is less indicated than 
Ar-Ar(Ca-Ca) collisions for photon-photon fusion processes with a typical
center of mass energy of ${\cal O}(100)$ GeV. Therefore,
the Pb-Pb mode will not be considered form this point on in our analysis.

\begin{figure}
\protect
\centerline{
\mbox{\epsfig{file=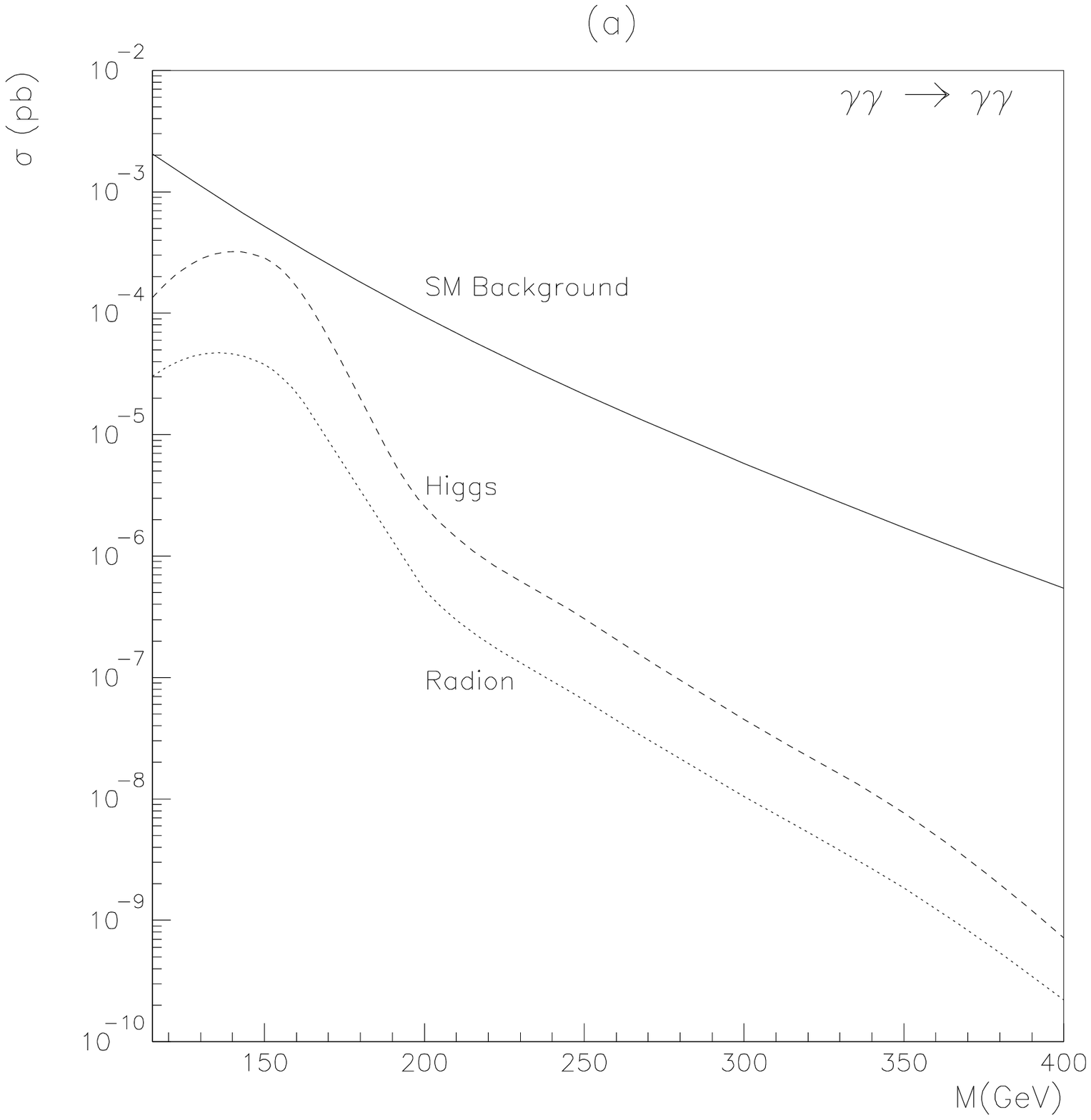,angle=0,width=0.6\textwidth}
      \epsfig{file=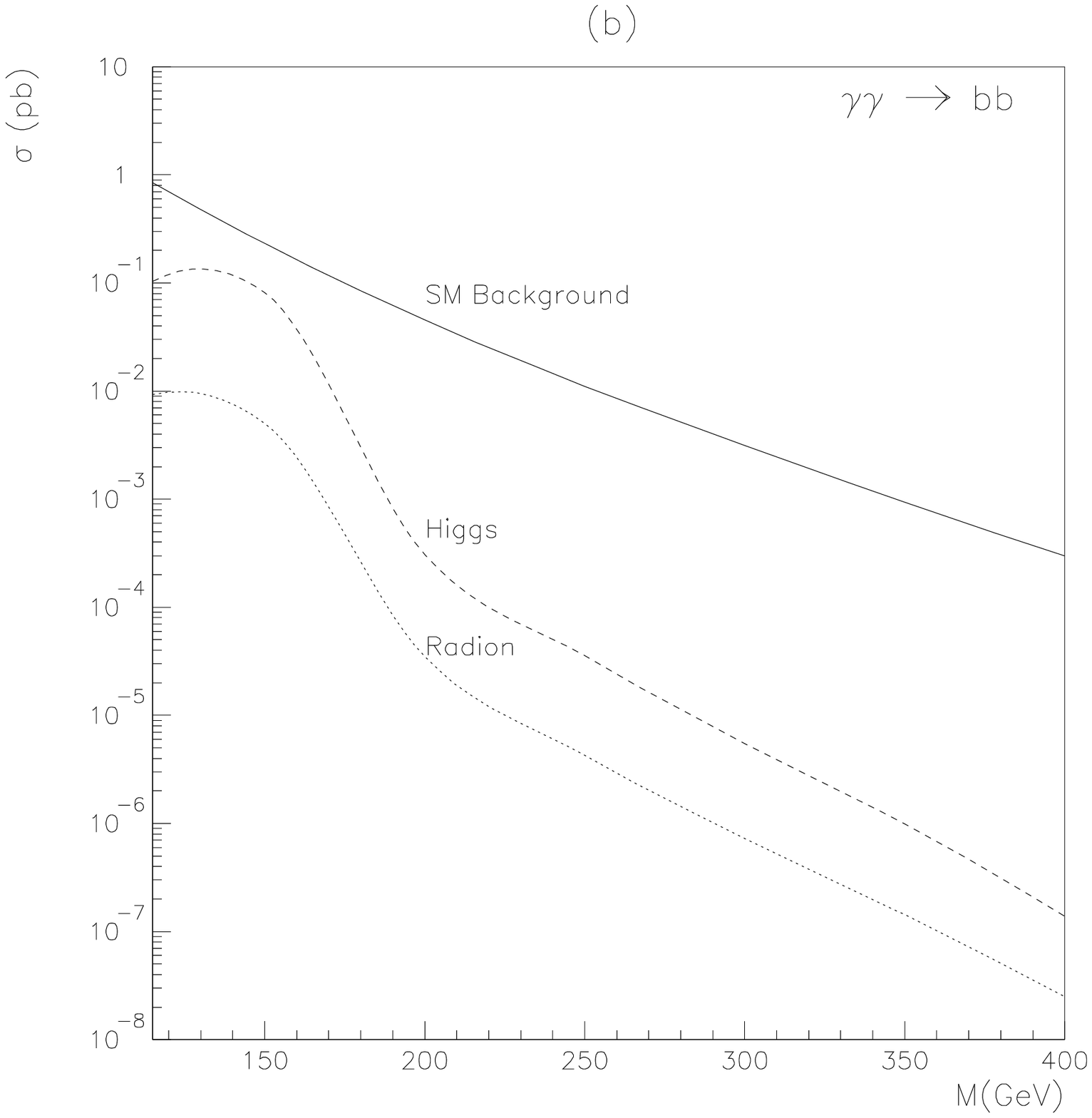,angle=0,width=0.6\textwidth}}}
\centerline{
\mbox{\epsfig{file=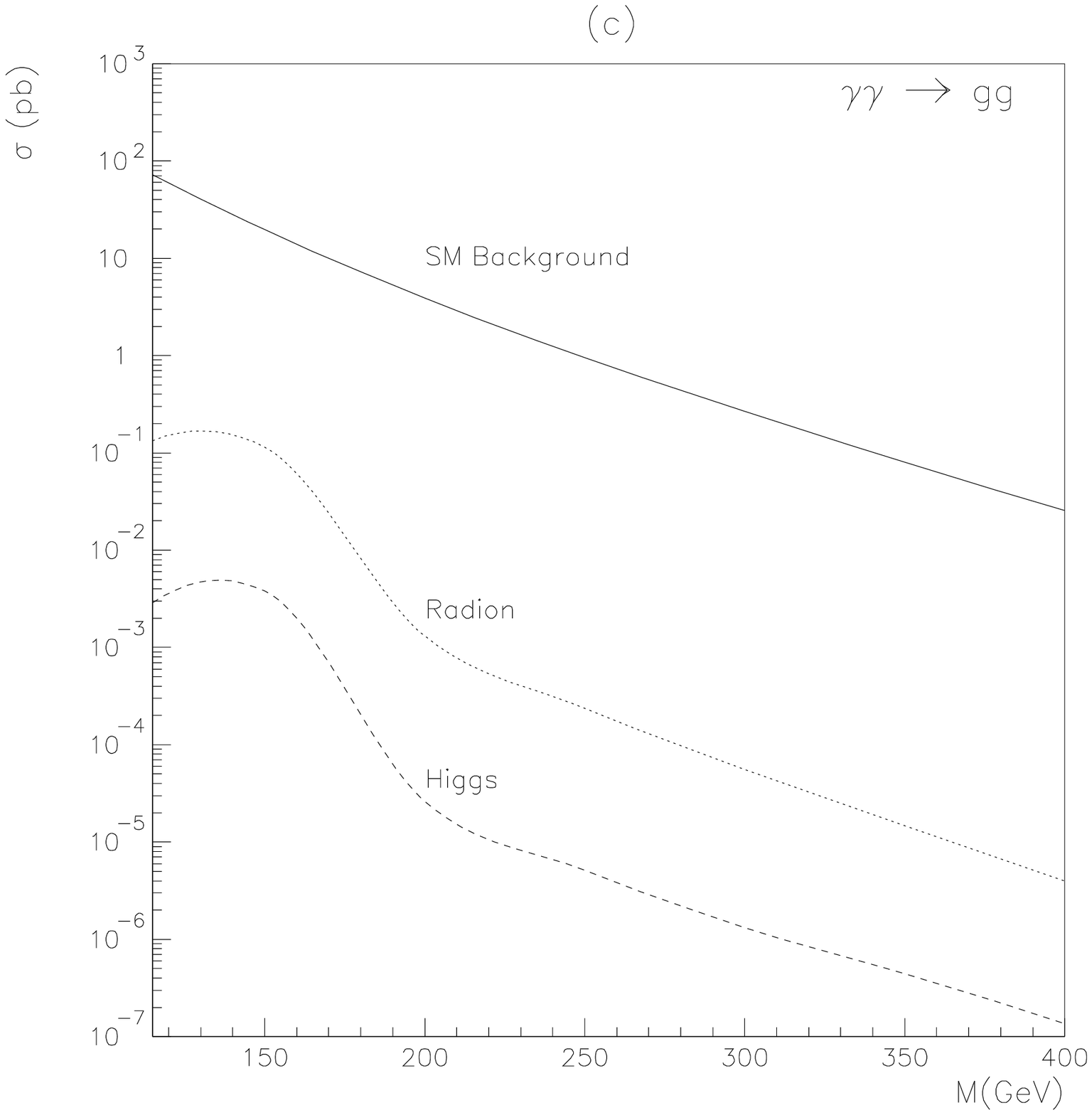,angle=0,width=0.6\textwidth}}}
\caption{Cross sections for the processes (a) $\gamma \gamma 
\to \gamma \gamma$, (b) $\gamma \gamma \to b \bar{b}$, and 
(c) $\gamma \gamma \to g g$, for $\lambda_R = 4v \approx 1$ TeV, 
considering events whose invariant masses fall in bins of size of 30 GeV 
around the mass $M$, as in Equation (\ref{cut2}). The full line
corresponds to the SM background discussed in the text while the 
dashed (dotted) line corresponds to the Higgs (Radion) contribution.}
\label{mass}
\end{figure}

Another information that can be extracted from Figure \ref{gg_lumi}
is the dependence between the effective photon-photon luminosity 
and the invariant mass of the initial $\gamma \gamma$ pair, 
which indicates that a discovery of a SM Higgs or a RSM Radion production
via photon-photon fusion is favored for low values for the invariant 
mass of the initial $\gamma \gamma$ pair. Similar conclusion can be 
obtained from Figure \ref{mass}, where the behavior of the cross sections 
of the processes  $\gamma \gamma \to \gamma \gamma$, 
$\gamma \gamma \to b \bar{b}$, and $\gamma \gamma \to g g$, 
for events whose invariant masses fall in bins of size of 30 GeV around 
the mass $M$ used to impose the cut in Equation (\ref{cut2}), is presented 
for $\lambda_R = 4v \approx 1$ TeV. Higher values for the cross sections 
are obtained for masses $M$ lower than 200 GeV. Therefore, from this point 
on, we will only consider in our analysis a SM Higgs and a RSM Radion
mass of 115 GeV, as indicated by the  latest hints from the 
LEP Higgs search \cite{Higgs} experiment.

\begin{figure}
\protect
\centerline{
\mbox{\epsfig{file=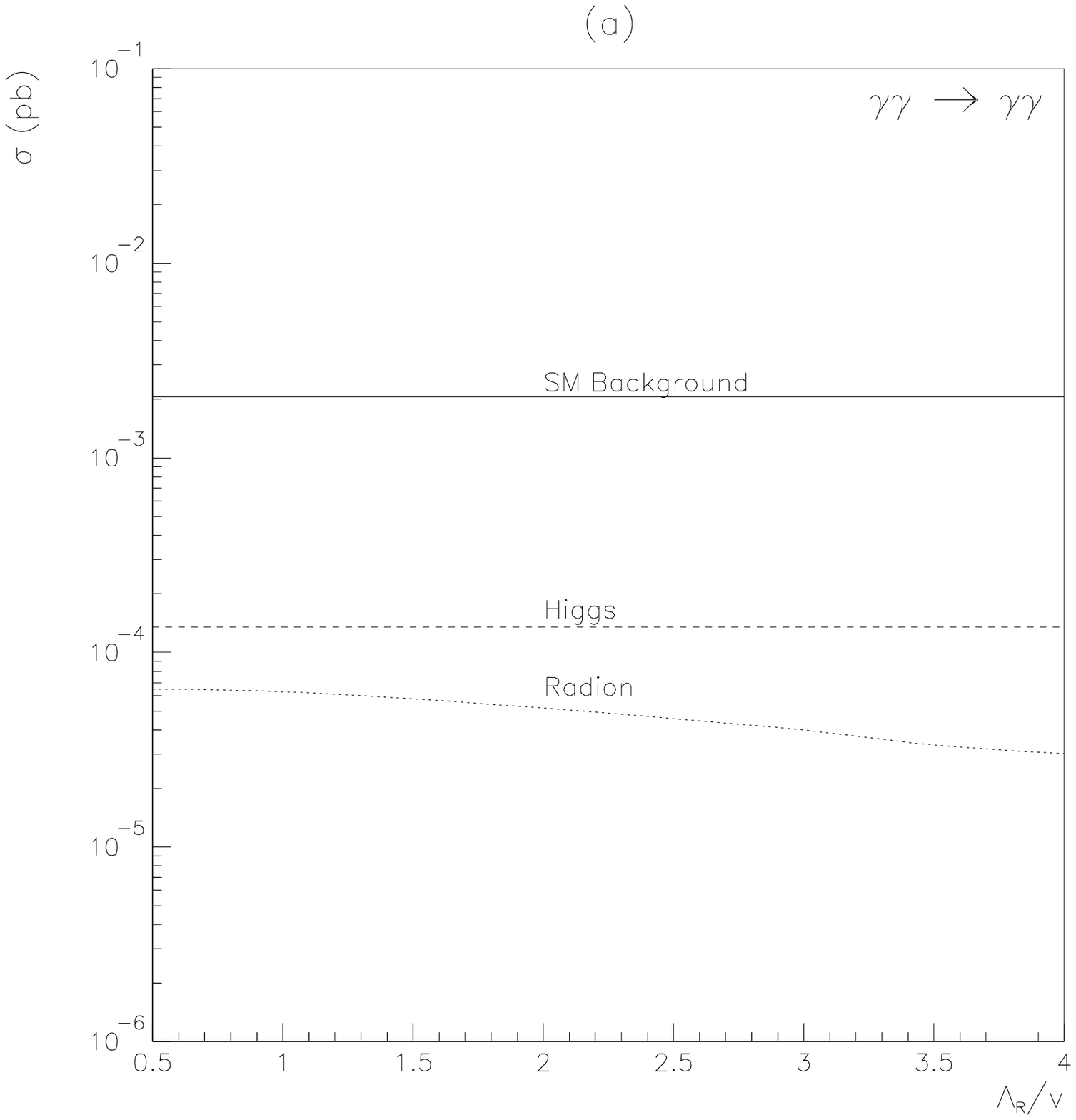,angle=0,width=0.6\textwidth}
      \epsfig{file=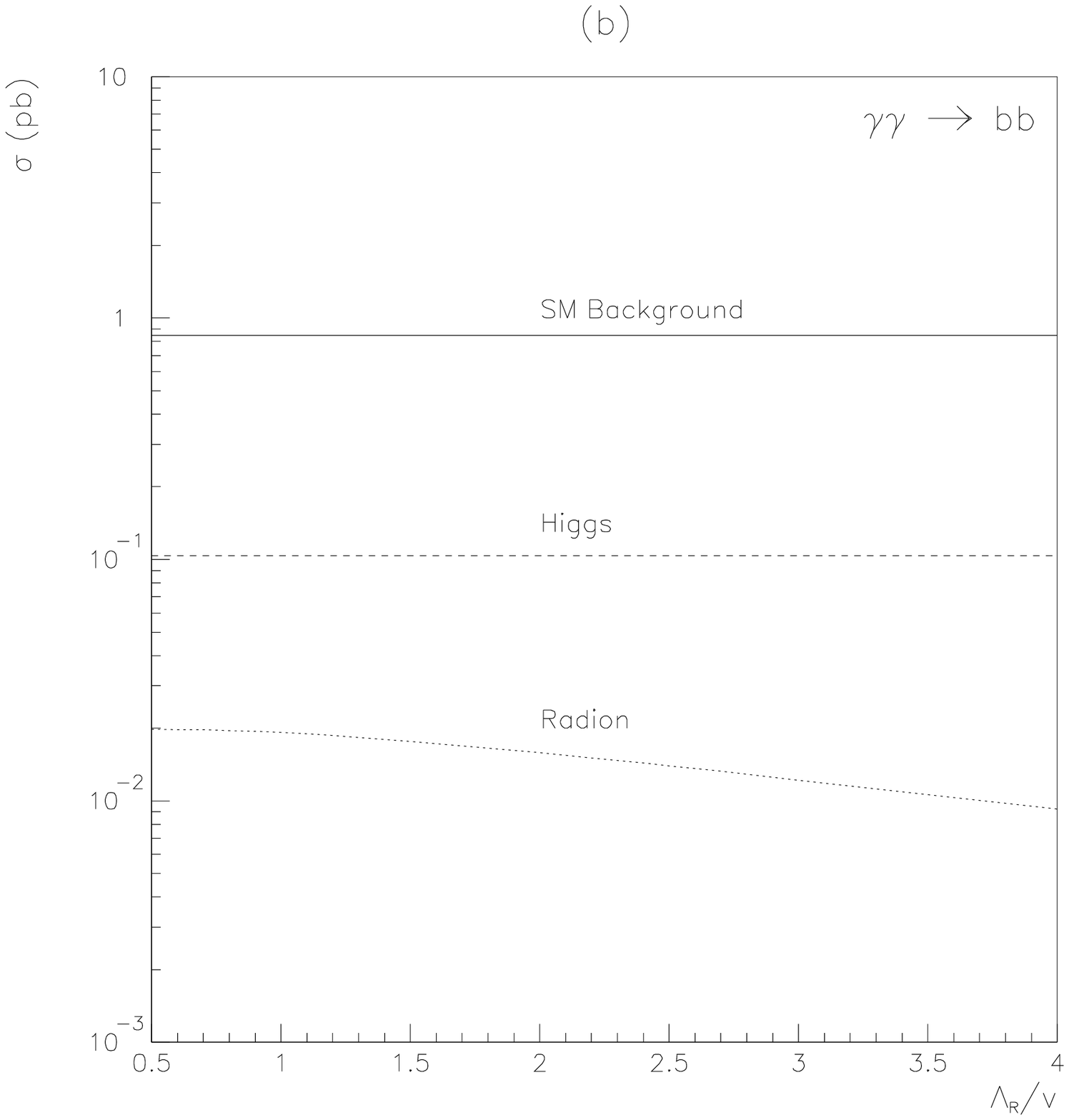,angle=0,width=0.6\textwidth}}}
\centerline{
\mbox{\epsfig{file=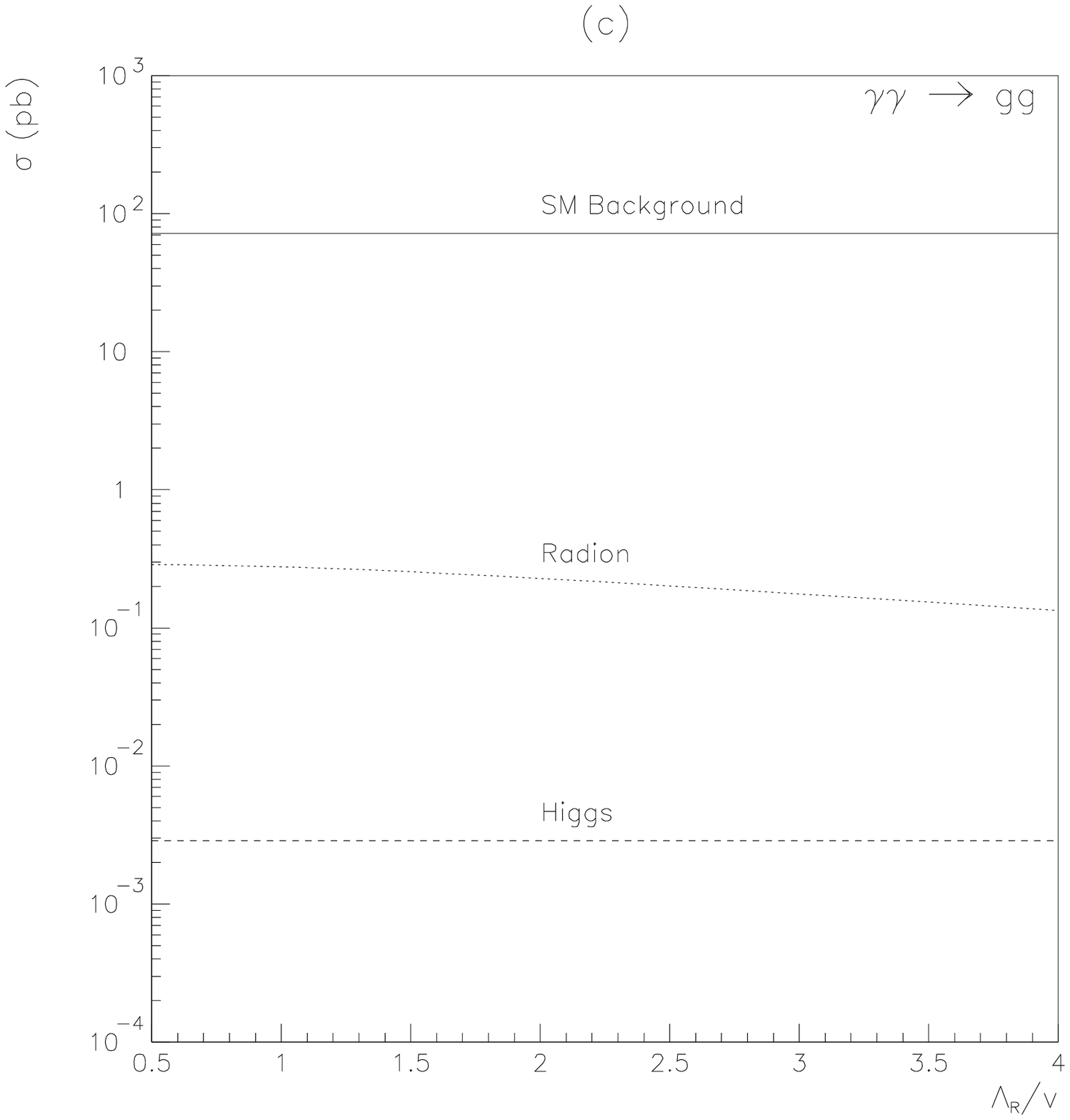,angle=0,width=0.6\textwidth}}}
\caption{Cross sections for the processes (a) $\gamma \gamma 
\to \gamma \gamma$, (b) $\gamma \gamma \to b \bar{b}$, and 
(c) $\gamma \gamma \to g g$ in terms of the ratio of the vev's of
the Radion ($\Lambda_R$) and the Higgs ($v$) fields. The mass of the
Higgs and/or Radion is equal to 115 GeV and  the set of cuts given by Equations
(\ref{cut1}) and (\ref{cut2}) was applied. 
The full line corresponds to 
the SM background discussed in the text while the dashed (dotted) line
corresponds to the Higgs (Radion) contribution.}
\label{lambda_v}
\end{figure}

Another point considered in our analyses is the dependence between the
cross sections for the Radion contribution of the three processes 
and the ratio of the vev's of the Radion ($\Lambda_R$) 
and the Higgs ($v$) fields. Figure \ref{lambda_v} shows the behavior of the
cross sections in the range $0.5 \leq \frac{\Lambda_R}{v} \leq 4$. 
Note that Figures \ref{lambda_v} (a) and (b) show that the SM Higgs
contribution is greater than the RSM Radion contribution in the processes 
$\gamma \gamma \to \gamma \gamma$ and $\gamma \gamma \to b \bar{b}$,
while Figure \ref{lambda_v} (c) shows that the RSM Radion contribution is
greater than the SM Higgs contribution in the process $\gamma \gamma \to gg$.
Therefore, the process $\gamma \gamma \to gg$ is the most sensitive for a 
Radion search while the other two processes are most sensitive for a Higgs 
search.

In order to identify a 95\% C.L. signal of a SM Higgs or a RSM Radion 
production at the heavy ion mode of the LHC, let us consider the 
significance ($S$) of a signal given by the equation
\begin{eqnarray}
 S = \frac{N_{Total} - N_{Background}}{\sqrt{N_{Total}}} = 
\frac{\sigma_{Total} - \sigma_{Background}}{\sqrt{\sigma_{Total}}}
\sqrt{\cal L},
\label{significance}
\end{eqnarray}
where $N$ is the number of events, $\cal L$ is the integrated luminosity
of the accelerator, $\sigma$ is cross section of the process considered.
The subscript $Background$ stands for the SM background contribution 
without any Higgs and/or Radion diagrams, and the subscript $Total$ 
stands for the total contribution, including Higgs and/or Radion diagrams. 
A 95\% C.L. signal is obtained when $S=1.96$ for Gaussian distributions. 
The results presented in Table \ref{refined_cut} for the 
$^{40}_{18}$Ar mode show that the SM background cross sections  
are at least one order of magnitude higher than the Higgs or Radion
signals. Note that if one has one event identified as a Higgs or Radion
exchange, than there will be at least ten SM background events, 
fact that justifies a Gaussian distribution approach.

Therefore, it is possible to evaluate the integrated luminosity needed for a
95\% C.L. Higgs signal by taking in Equation (\ref{significance}) $S=1.96$
and the cross sections presented in Table \ref{refined_cut}, 
with $\sigma_{Total}$ given by the sum $(\sigma_{Background} + 
\sigma_{Higgs})$ since the interference effects are negligible, as checked
in our MadGraph/Helas code. The results are presented in Table 
\ref{lumi_Higgs_cut}, where the number of years needed to establish a 
95\% C.L. Higgs signal is also shown when we consider the accelerator 
luminosity given by ${\cal L} = 0.63$ pb$^{-1}$ year$^{-1}$, as 
discussed above in the text. Table  \ref{lumi_Higgs_cut} also shows the 
results for the Ca-Ca mode of the accelerator, with luminosity given by 
${\cal L} = 6$ pb$^{-1}$ year$^{-1}$. Analogously, a 95\% C.L. Higgs plus 
Radion signal can be considered by simply taking $\sigma_{Total} = 
(\sigma_{Background} + \sigma_{Higgs} + \sigma_{Radion})$, and the results 
for this case is presented in Table \ref{lumi_Higgs_Radion_cut}.

The results in Table \ref{lumi_Higgs_cut} indicate that the process 
$\gamma \gamma \to b \bar{b}$ is the best choice to search the Higgs 
boson because the integrated luminosity needed for a 95\% C.L. 
signal ($\approx 250$ pb$^{-1}$) is three orders of magnitude smaller 
than the luminosity needed for the process $\gamma \gamma \to 
\gamma \gamma$ and five  orders of magnitude smaller than the luminosity 
needed for the process $\gamma \gamma \to gg$. However, this integrated 
luminosity is still very high compared to the luminosity expected for 
both Ar-Ar and Ca-Ca mode, tens of years being needed for a 95\% C.L.  
signal detection.  

The results in Table \ref{lumi_Higgs_Radion_cut} include the RMS Radion
in the analysis. There are small changes for the $\gamma \gamma
\to b \bar{b}$ and $\gamma \gamma \to \gamma \gamma$ processes. The main 
difference appears in the $\gamma \gamma \to g g$ process, where the 
integrated luminosity needed for a 95\% C.L.  signal is three orders of 
magnitude smaller than the one of Table  \ref{lumi_Higgs_cut}. The reason
for this change is that the Radion contribution is greater than the Higgs
contribution only in the process $\gamma \gamma \to g g$, as can be seen 
in Figs. \ref{mass}(c) and \ref{lambda_v}(c). 

In order to improve the results, one could collected final states 
$\gamma \gamma$, $b \bar{b}$ and $gg$ events whose invariant masses 
fall in bins of size of 10 GeV around the Higgs (Radion) mass
\begin{eqnarray}
m_{H(R)} - 5\text{ GeV}
< m_{\gamma \gamma (b \bar{b}) [gg]}  < m_{H(R)} + 5\text{ GeV}.
\;\;
\label{cut3}
\end{eqnarray}
In this situation, the SM background is even more reduced while the Higgs
and Radion signals are unchanged, as can be seen in
Table \ref{super_refined_cut}. The total integrated luminosity needed
for a 95\% C.L. Higgs (Higgs plus Radion) signal are now presented in 
Table \ref{lumi_Higgs_supercut} (\ref{lumi_Higgs_Radion_supercut}).

The number of years needed for a 95\% C.\ L.\ Higgs signal in
the process $\gamma \gamma \to b \bar{b}$ at the Ca-Ca mode of the
accelerator is reduced to $\approx$ 15 years. When the Radion is included 
in the analysis, the number of years is reduced to  $\approx$ 12.5 years. 
If the experiment luminosity could be enhanced by a factor of ten,
then a 95\% C.\ L.\ SM Higgs signal could be obtained in 18 months. Still 
in this case, if a  95\% C.\ L.\ signal were obtained in 15 months, it
would be an indication of the existence of the RMS Radion.  
If the experiment luminosity could be enhanced by a factor of twenty,
then a 95\% C.\ L.\ RMS Radion signal in the process $\gamma \gamma \to
gg$ could be obtained in 31 months.

\section{Conclusions}
\label{sec:con}

In this work we have studied the sensitivity of the heavy ion mode of
the LHC to detect the production of Higgs and Radion scalars via 
photon-photon fusion through the analysis of the processes
$\gamma \gamma \to \gamma\gamma, b \bar{b}$ and $gg$
in peripheral heavy ion collisions.

The chances of finding  the SM Higgs boson (or the RMS Radion) are marginal
for high values of the Higgs (Radion) mass. For lower masses the situation
is still critical, but there is some hope left. We have considered 
$M_H=M_R=115$ GeV in our analysis according to the recent LEP hints on
the Higgs mass.     

The best place to search the Higgs boson is in the Ca-Ca ion mode 
of the LHC accelerator through the analysis of the process
$\gamma \gamma \to b \bar{b}$. In this case, considering the luminosities
presented in the literature, a 95\% C.\ L.\ signal can be established in
15 years of run. If the Radion scalar of the RSM is taken into account, a  
95\% C.\ L.\ signal would  be established  in 12.5 years. If the 
experiments could enhance their expectation for the luminosity 
by a factor of 10, then a  95\% C.\ L.\ SM Higgs signal could be 
established in less than two years of run.

On the other hand, the best place to  search the Radion of the 
RSM is in the Ca-Ca ion mode of the LHC accelerator through the analysis 
of the process $\gamma \gamma \to gg$. In this case, the experiments 
would have to improve their luminosity prediction by a factor of twenty 
in order to establish a 95\% C.\ L.\ Radion  signal in less than three
years of run.

In conclusion, SM Higgs and RMS Radion observation in the heavy ion mode 
of the LHC accelerator is improbable, unless the expected luminosity of
the  experiment could be enhanced by a factor of 10--20.

\acknowledgments

S.\ M.\ L.\ wishes to thank O.\ J.\ P.\ \'Eboli for encouragement and 
C.\ G.\ R.\ wishes to thank S.\ Klein for useful discussions. This work 
was supported by Funda\c{c}\~ao de Amparo \`a Pesquisa do Estado de 
S\~ao Paulo (FAPESP).


\widetext

\begin{table}
\begin{tabular}{||c||c||c||c||c||}
Ion considered & Final State & $\sigma_{Background}$ (pb) &
$\sigma_{Higgs}$ (pb) & $\sigma_{Radion}$ (pb) \\
\hline
\hline
$^{40}_{18}$Ar & $\gamma \gamma$ & 1.961$\times 10^{-2}$  & 1.346$\times 10^{-4}$ & 3.020$\times 10^{-5}$ \\
\hline
$^{40}_{18}$Ar & $b \bar{b}$     & 6.927$\times 10^{0}$   & 1.038$\times 10^{-1}$ & 8.982$\times 10^{-3}$ \\
\hline
$^{40}_{18}$Ar & $gg$            & 5.682$\times 10^{2}$   & 2.874$\times 10^{-3}$ & 1.334$\times 10^{-1}$ \\
\hline
\hline
$^{208}_{82}$Pb & $\gamma \gamma$ & 1.160$\times 10^{0}$   & 3.913$\times 10^{-3}$ & 8.627$\times 10^{-4}$ \\
\hline
$^{208}_{82}$Pb & $b \bar{b}$     & 3.919$\times 10^{2}$   & 3.023$\times 10^{0}$  & 2.666$\times 10^{-1}$ \\
\hline
$^{208}_{82}$Pb & $gg$            & 3.179$\times 10^{4}$   & 8.344$\times 10^{-2}$ & 3.883$\times 10^{0}$
\end{tabular}
\medskip
\caption{Cross Section in pb for the process $\gamma \gamma \to$ Final State
with $m_H=m_R=115$ GeV and $\Lambda_R = 4v = 984$ GeV $\approx 1$ TeV in
heavy ion collisions at LHC with the acceptance set of cuts of
Equation (\ref{cut1}). $\sigma_{Background}$ stands for the SM background,
$\sigma_{Higgs}$ stands for the contribution $\gamma \gamma \to H \to$
Final State and $\sigma_{Radion}$ stands for the contribution 
$\gamma \gamma \to R \to$ Final State.}
\label{acceptance_cut}
\end{table}

\begin{table}
\begin{tabular}{||c||c||c||c||c||}
Ion considered & Final State & $\sigma_{Background}$ (pb) &
$\sigma_{Higgs}$ (pb) & $\sigma_{Radion}$ (pb) \\
\hline
\hline
$^{40}_{18}$Ar & $\gamma \gamma$ & 2.050$\times 10^{-3}$  & 1.349$\times 10^{-4}$ & 3.030$\times 10^{-5}$ \\
\hline
$^{40}_{18}$Ar & $b \bar{b}$     & 8.486$\times 10^{-1}$  & 1.038$\times 10^{-1}$ & 9.254$\times 10^{-3}$ \\
\hline
$^{40}_{18}$Ar & $gg$            & 7.170$\times 10^{1}$   & 2.875$\times 10^{-3}$ & 1.338$\times 10^{-1}$ \\
\hline
\hline
$^{208}_{82}$Pb & $\gamma \gamma$ & 6.589$\times 10^{-2}$  & 4.103$\times 10^{-3}$ & 9.222$\times 10^{-4}$ \\
\hline
$^{208}_{82}$Pb & $b \bar{b}$     & 2.721$\times 10^{1}$   & 3.159$\times 10^{0}$  & 2.818$\times 10^{-1}$ \\
\hline
$^{208}_{82}$Pb & $gg$            & 2.298$\times 10^{3}$   & 8.753$\times 10^{-2}$ & 4.072$\times 10^{0}$
\end{tabular}
\medskip
\caption{Cross Section in pb for the process $\gamma \gamma \to$ Final State
with $m_H=m_R=115$ GeV and $\Lambda_R = 4v = 984$ GeV $\approx 1$ TeV in
heavy ion collisions at LHC with the refined set of cuts of
Equations (\ref{cut1}) and  (\ref{cut2}). $\sigma_{Background}$ stands 
for the SM background without Higgs and/or Radion diagrams,
$\sigma_{Higgs}$ stands for the contribution $\gamma \gamma \to H \to$
Final State and $\sigma_{Radion}$ stands for the contribution 
$\gamma \gamma \to R \to$ Final State.}
\label{refined_cut}
\end{table}

\begin{table}
\begin{tabular}{||c||c||c||c||}
Ion considered & Final State & Integrated Luminosity (pb$^{-1}$) &
Years \\
\hline
\hline
$^{40}_{18}$Ar &$\gamma \gamma$ &4.612$\times 10^{5}$ &7.321$\times 10^{5}$ \\
\hline
$^{40}_{18}$Ar &$b \bar{b}$     &3.396$\times 10^{2}$ &5.390$\times 10^{2}$ \\
\hline
$^{40}_{18}$Ar &$gg$            &3.333$\times 10^{7}$ &5.290$\times 10^{7}$ \\
\hline
\hline
$^{40}_{20}$Ca & $\gamma \gamma$&3.026$\times 10^{5}$ &5.044$\times 10^{4}$ \\
\hline
$^{40}_{20}$Ca & $b \bar{b}$    &2.228$\times 10^{2}$ &3.713$\times 10^{1}$ \\
\hline
$^{40}_{20}$Ca & $gg$           &2.186$\times 10^{7}$ &3.644$\times 10^{6}$ 
\end{tabular}
\medskip
\caption{Total Integrated Luminosity needed for a 95\% C.L. Higgs signal 
for the process $\gamma \gamma \to$ Final State with $m_H=115$ GeV 
in heavy ion collisions at LHC with the refined set of cuts of Equations 
(\ref{cut1}) and (\ref{cut2}). It is also presented the number of years 
needed for a 95\% C.L. Higgs signal considering a luminosity of 
0.63(6)pb$^{-1}$ year$^{-1}$ for the Ar-Ar (Ca-Ca) mode as 
discussed in the text.}
\label{lumi_Higgs_cut}
\end{table}

\begin{table}
\begin{tabular}{||c||c||c||c||}
Ion considered & Final State & Integrated Luminosity (pb$^{-1}$) &
Years \\
\hline
\hline
$^{40}_{18}$Ar &$\gamma \gamma$ &3.118$\times 10^{5}$ &4.950$\times 10^{5}$ \\
\hline
$^{40}_{18}$Ar &$b \bar{b}$     &2.890$\times 10^{2}$ &4.588$\times 10^{2}$ \\
\hline
$^{40}_{18}$Ar &$gg$            &1.477$\times 10^{4}$ &2.345$\times 10^{4}$ \\
\hline
\hline
$^{40}_{20}$Ca & $\gamma \gamma$&2.046$\times 10^{5}$ &3.410$\times 10^{4}$ \\
\hline
$^{40}_{20}$Ca & $b \bar{b}$    &1.896$\times 10^{2}$ &3.161$\times 10^{1}$ \\
\hline
$^{40}_{20}$Ca & $gg$           &9.693$\times 10^{3}$ &1.615$\times 10^{3}$ 
\end{tabular}
\medskip
\caption{Total Integrated Luminosity needed for a 95\% C.L. Higgs plus 
Radion signal for the process $\gamma \gamma \to$ Final State with 
$m_H=m_R=115$ GeV and $\Lambda_R = 4v = 984$ GeV $\approx 1$ TeV 
in heavy ion collisions at 
LHC with the refined set of cuts of Equations (\ref{cut1}) and (\ref{cut2}). 
It is also presented the number of years needed for a 95\% C.L. Higgs plus 
Radion signal considering a luminosity of 0.63(6)pb$^{-1}$ year$^{-1}$ for 
the Ar-Ar (Ca-Ca) mode as discussed in the text.}
\label{lumi_Higgs_Radion_cut}
\end{table}

\begin{table}
\begin{tabular}{||c||c||c||c||c||}
Ion considered & Final State & $\sigma_{Background}$ (pb) &
$\sigma_{Higgs}$ (pb) & $\sigma_{Radion}$ (pb) \\
\hline
\hline
$^{40}_{18}$Ar & $\gamma \gamma$ & 6.433$\times 10^{-4}$  & 1.346$\times 10^{-4}$ & 3.024$\times 10^{-5}$ \\
\hline
$^{40}_{18}$Ar & $b \bar{b}$     & 2.682$\times 10^{-1}$  & 1.036$\times 10^{-1}$ & 9.252$\times 10^{-3}$ \\
\hline
$^{40}_{18}$Ar & $gg$            & 2.268$\times 10^{1}$   & 2.874$\times 10^{-3}$ & 1.334$\times 10^{-1}$ 
\end{tabular}
\medskip
\caption{Cross Section in pb for the process $\gamma \gamma \to$ Final State
with $m_H=m_R=115$ GeV and $\Lambda_R = 4v = 984$ GeV $\approx 1$ TeV in
heavy ion collisions at LHC with the refined set of cuts of
Equations (\ref{cut1}) and  (\ref{cut3}). $\sigma_{Background}$ stands 
for the SM background without Higgs and/or Radion diagrams,
$\sigma_{Higgs}$ stands for the contribution $\gamma \gamma \to H \to$
Final State and $\sigma_{Radion}$ stands for the contribution 
$\gamma \gamma \to R \to$ Final State.}
\label{super_refined_cut}
\end{table}

\begin{table}
\begin{tabular}{||c||c||c||c||}
Ion considered & Final State & Integrated Luminosity (pb$^{-1}$) &
Years \\
\hline
\hline
$^{40}_{18}$Ar &$\gamma \gamma$ &1.649$\times 10^{5}$ &2.618$\times 10^{5}$ \\
\hline
$^{40}_{18}$Ar &$b \bar{b}$     &1.331$\times 10^{2}$ &2.112$\times 10^{2}$ \\
\hline
$^{40}_{18}$Ar &$gg$            &1.055$\times 10^{7}$ &1.675$\times 10^{7}$ \\
\hline
\hline
$^{40}_{20}$Ca & $\gamma \gamma$&1.082$\times 10^{5}$ &1.804$\times 10^{4}$ \\
\hline
$^{40}_{20}$Ca & $b \bar{b}$    &8.731$\times 10^{1}$ &1.455$\times 10^{1}$ \\
\hline
$^{40}_{20}$Ca & $gg$           &6.922$\times 10^{6}$ &1.154$\times 10^{6}$ 
\end{tabular}
\medskip
\caption{Total Integrated Luminosity needed for a 95\% C.L. Higgs signal 
for the process $\gamma \gamma \to$ Final State with $m_H=115$ GeV 
in heavy ion collisions at LHC with the refined set of cuts of Equations 
(\ref{cut1}) and (\ref{cut3}). It is also presented the number of years 
needed for a 95\% C.L. Higgs signal considering a luminosity of 
0.63(6)pb$^{-1}$ year$^{-1}$ for the Ar-Ar (Ca-Ca) mode as 
discussed in the text.}
\label{lumi_Higgs_supercut}
\end{table}

\begin{table}
\begin{tabular}{||c||c||c||c||}
Ion considered & Final State & Integrated Luminosity (pb$^{-1}$) &
Years \\
\hline
\hline
$^{40}_{18}$Ar &$\gamma \gamma$ &1.143$\times 10^{5}$ &1.814$\times 10^{5}$ \\
\hline
$^{40}_{18}$Ar &$b \bar{b}$     &1.149$\times 10^{2}$ &1.824$\times 10^{2}$ \\
\hline
$^{40}_{18}$Ar &$gg$            &4.720$\times 10^{3}$ &7.492$\times 10^{3}$ \\
\hline
\hline
$^{40}_{20}$Ca & $\gamma \gamma$&7.496$\times 10^{4}$ &1.249$\times 10^{4}$ \\
\hline
$^{40}_{20}$Ca & $b \bar{b}$    &7.541$\times 10^{1}$ &1.257$\times 10^{1}$ \\
\hline
$^{40}_{20}$Ca & $gg$           &3.097$\times 10^{3}$ &5.161$\times 10^{2}$ 
\end{tabular}
\medskip
\caption{Total Integrated Luminosity needed for a 95\% C.L. Higgs plus 
Radion signal for the process $\gamma \gamma \to$ Final State with 
$m_H=m_R=115$ GeV and $\Lambda_R = 4v = 984$ GeV $\approx 1$ TeV 
in heavy ion collisions at 
LHC with the refined set of cuts of Equations (\ref{cut1}) and (\ref{cut3}). 
It is also presented the number of years needed for a 95\% C.L. Higgs plus 
Radion signal considering a luminosity of 0.63(6)pb$^{-1}$ year$^{-1}$ for 
the Ar-Ar (Ca-Ca) mode as discussed in the text.}
\label{lumi_Higgs_Radion_supercut}
\end{table}

\end{document}